\documentclass{elsart}

\usepackage{amssymb,bm,amsmath}

\newcommand{\be}{\begin{equation}}
\newcommand{\ee}{\end{equation}}
\newcommand{\ba}{\begin{eqnarray}}
\newcommand{\ea}{\end{eqnarray}}

\newcommand{\nn}{\nonumber}

\newcommand{\BV}{\mbox{$\langle 0 |$}}
\newcommand{\KV}{\mbox{$| 0 \rangle$}}
\newcommand{\bra}[1]{\mbox{$\langle #1|$}}
\newcommand{\ket}[1]{\mbox{$| #1\rangle$}}
\newcommand{\braket}[2]{\mbox{$\langle #1| #2\rangle $}}

\newcommand{\BBV}{\mbox{$\langle \! \langle 0 |$}}
\newcommand{\KKV}{\mbox{$| 0 \rangle \! \rangle$}}
\newcommand{\bbra}[1]{\mbox{$\langle \! \langle #1|$}}
\newcommand{\kket}[1]{\mbox{$| #1 \rangle \! \rangle$}}

\newcommand{\rme}{{\rm e}}
\newcommand{\rmi}{{\rm i}}

\newcommand{\bmx}{\bm{x}}

\begin{document}

\begin{frontmatter}



\title{Quantum Field Theoretical Description of Unstable Behavior of
Trapped Bose-Einstein Condensates with Complex
Eigenvalues of Bogoliubov-de Gennes Equations} 


\author[MM]{Makoto Mine},
\ead{mine@aoni.waseda.jp}
\author[MO]{Masahiko Okumura},
\ead{okumura@aoni.waseda.jp}
\author[MM]{Tomoka Sunaga} and
\ead{tomoka@fuji.waseda.jp}
\author[YY]{Yoshiya Yamanaka}
\ead{yamanaka@waseda.jp}

\address[MM]{Department of Physics, Waseda University, Tokyo 169-8555,
 Japan}
\address[MO]{Department of Applied Physics, Waseda University, Tokyo
 169-8555, Japan}
\address[YY]{Department of Materials Science and Engineering, Waseda
 University, \\ Tokyo 169-8555, Japan}

\begin{abstract}
The Bogoliubov-de Gennes equations are used for a number of theoretical
 works on the trapped Bose-Einstein condensates. 
These equations are known to give the energies of the quasi-particles
 when all the eigenvalues are real.
We consider the case in which these equations have complex eigenvalues.
We give the complete set including those modes whose eigenvalues
are complex. The quantum fields which represent neutral atoms are
expanded in terms of the complete set.  It is shown that the state
space is an indefinite metric one and that the free Hamiltonian is not
diagonalizable in the conventional bosonic representation. 
We introduce a criterion to select quantum states describing the
metastablity of the condensate,
called the physical state conditions.
In order to study the instability, we formulate the linear response of
the density against the time-dependent external perturbation within the
regime of Kubo's linear response theory. 
Some states, satisfying all the physical state conditions, give the
blow-up and damping behavior of the density distributions corresponding
to the complex eigenmodes. 
It is qualitatively consistent with the result of the  recent analyses using the
time-dependent Gross-Pitaevskii equation. 
\end{abstract}

\end{frontmatter}

\section{\label{intro}Introduction}

In 1995, the Bose-Einstein condensates (BECs) of neutral atoms in the
magnetic trap were realized \cite{Boulder,MIT,Li}. After several years,
some kinds of quantized vortices in condensates have been observed,
e.g., singly quantized vortices and a vortex 
lattice \cite{Matthews,Madison,Abo-Shaeer}.
Furthermore, the doubly quantized vortices have been created under the
use of phase imprinting technique \cite{Leanhardt}. 
It has also been observed that a doubly quantized vortex decays into
two singly quantized vortices \cite{Shin}.

Excitations around the condensates, with or without vortices, are
represented by the solutions of the Bogoliubov-de Gennes (BdG) 
equations \cite{Bogoliubov,de Gennes,Fetter}, which 
follow from linearizing the
time-dependent Gross-Pitaevskii (TDGP) equation \cite{PandS} and
form a set of two-component eigenfunctions. 
If all the engenvalues are real, 
the corresponding free Hamiltonian is diagonalized in the bosonic
representation and the quasi-particle picture is obtained.

Recently, it is found that the BdG equations have complex eigenvalues in
some cases, e.g., the case where the condensate flows in an optical
lattice \cite{Wu}, or where the condensate has a highly quantized 
vortex \cite{Pu,Garay,Skryabin,Mettenen,Kawaguchi} 
or gap solitons \cite{Hilligsoe}, 
or the case of the multi-component BECs \cite{Zhang,Robert}.
It is also mentioned that a subject of the complex eigenvalues is taken
up in the context of diagonalization of the Hamiltonian of
the quadratic form of creation and annihilation operators \cite{RK} 
and in the studies of the Gross-Pitaevskii (GP) energy 
functional \cite{GPstability}. 
These eigenfunctions with complex eigenvalues cause the ``dynamical
instability'', which is a type of instability of the condensates and should
be distinguished from the ``Landau instability'' caused by negative energies
of the quasi-particle. 
The former is associated with the decay of the initial configuration of
the condensate, 
while the latter corresponds to the absence of thermodynamic 
equilibrium \cite{PandS}.
The dynamical instability of the condensates in optical lattices
was observed in the experiment, and the values of the quasi-momentum at
which the experimental loss rate has its peak are in good agreement with
those at which the imaginary part of the complex eigenvalue 
of the BdG equations has its peak \cite{LatticeExp}.
The time scale to form the ferromagnetic
domains in the spinor BECs, observed
experimentally by rapidly quenching the system to conditions in which
the ferromagnetic phase is energetically favored, 
is roughly described by the imaginary part of the complex eigenvalue
of the BdG equations \cite{Sadler}.
In addition, when
the condensate has a highly quantized vortex, 
the emergence of the complex eigenvalues is interpreted as the sign of
dynamical instability caused by the external 
perturbations \cite{Mettenen,Kawaguchi}. 
However, no corresponding experiment in which the condensate is perturbed by an
external force has been reported, although the decay of a doubly
quantized vortex into two singly quantized vortices, without
perturbation, is observed experimentally \cite{Shin}. 
The relation between the complex eigenvalues and the instability of highly
quantized vortices is not elucidated fully and is still under 
study \cite{Gawryluk,Huhtameki}.

So far, the interpretation of complex eigenvalues in the BdG equations
has been based on the TDGP equation, but not on the quantum field theory
(QFT).  
It has not been made clear how the complex eigenmodes should be
interpreted in QFT, and how to describe the dynamical
instability against external time-dependent perturbations within the
formulation of QFT. 
In general, it is possible in terms of equilibrium QFT to give dynamical
description in the linear regime, using Kubo's linear response theory
(LRT) \cite{LRTtext,Minguzzi,Kubo}. 
But Kubo's LRT is based on the canonical commutation relations (CCRs)
and representations of the states. 
In the case where complex eigenvalues arise, it is not easy to keep the
CCRs and to discuss the representations of states due to the indefinite
metric which is characteristic of the BdG equations. 
It is well known in QFT that when the indefinite metric is involved,
careful treatments on the complete set are needed
(for example, the {\it ghosts} in gauge fields \cite{Peskin}).
But this point has not been discussed in the framework of QFT with
complex eigenvalues of the BdG equations and the meaning of the complex
eigenmodes in Kubo's LRT has not been clear.

In this paper, we construct QFT and formulate Kubo's LRT in the case
where complex eigenvalues arise in the BdG equations. 
The blow-up or damping behavior of the density distributions, corresponding
to the complex modes against the time-dependent external perturbation, 
is shown in the linear regime.  
It is consistent with the result from the recent analyses based on the
TDGP equation \cite{Mettenen,Kawaguchi}. 
Furthermore, we discuss that the effect beyond the GP theory is given
if we consider the higher order terms of quantum fluctuations using the
formalism established in this paper. 

This paper is organized as follows.
In Sec.~\ref{sec-Model}, the model action and Hamiltonian are given. 
In Sec.~\ref{sec-BdG}, the BdG equations are introduced, and when 
all the eigenvalues are real, 
the mathematical properties of eigenfunctions including the zero-mode are
summarized. 
In Sec.~\ref{sec-doublet}, for the sake of convenience in the 
following discussions, the doublet notations are introduced. 
In Sec.~\ref{sec-complexBdG}, we consider the case in which the complex
eigenvalues arise in the BdG equations. 
Mathematical properties of the eigenfunctions are summarized and
the complete set of functions is given. 
In Sec.~\ref{sec-Hamiltonian}, the quantized fields are expanded 
using the complete set of functions given in Sec.~\ref{sec-complexBdG}. 
We check that the CCRs are kept in this expansion. 
Next, we study the properties of the Hamiltonian. 
Particularly, we show that the free Hamiltonian can not be diagonalized
in the usual bosonic representation. 
Then we construct eigenstates of the free Hamiltonian, which are not
normalizable.  
We discuss the complete set of states, corresponding to the complex
eigenspace.  
In Sec.~\ref{sec-Phys}, 
we introduce the conditions for physical states,
which are not only matched with the result of the experiment in which 
highly quantized vortex states are metastable, but also provides us with
a consistent QFT description of the unstable behavior. 
We examine some candidates of states as to whether all the conditions are 
satisfied. 
In Sec.~\ref{sec-LRT}, for the physical states examined in
Sec.~\ref{sec-Phys}, using Kubo's LRT, we calculate the density response
of the system against the external perturbation of both impulsive and
periodic types. 
As a result, we show that the complex eigenvalues correspond to the
instability against some external time-dependent perturbations to the
condensates. 
Section \ref{sec-Sum} is devoted to summary.

\section{Model Action and Hamiltonian}\label{sec-Model}
We start with the following action to describe the trapped BEC of
neutral atoms,
\begin{equation}
S = \int \! d^4 x \left[
\psi^\dag (x) \{ T - K - V ({\bm x}) + \mu \} \psi (x)
- \frac{g}{2} \psi^{\dag 2} (x) \psi^2 (x) \right] \, , \label{S}
\end{equation}
where $x=(\bmx, t)$, and
\label{TKV}
\begin{eqnarray}
T &=& i \frac{\partial}{\partial t} \, , \\
K &=& - \frac{1}{2m} \nabla^2 \, , \\
V ({\bm x}) &=& \frac{1}{2} m ( \omega_x^2 x^2 + \omega_y^2 y^2 +
\omega_z^2 z^2) \, ,
\end{eqnarray}
with the mass of the neutral atoms $m$, the chemical potential $\mu$ and
the coupling constant $g$. Here we have written the trapping potential
of harmonic type, but the discussions in this paper are valid for more
general type of potentials. 
Throughout this paper $\hbar$ is set to be unity.

In the terminology of the canonical operator formalism, let us divide
the original field ${\psi}(x)$ into the classical and quantum parts as
\begin{equation}
{\psi}(x) = \zeta ({\bm x}) + {\varphi} (x) \, , \label{psi}
\end{equation}
where it is assumed that the c-number function $\zeta({\bm x})$, which is
the order parameter, is time-independent.
Note that the function $\zeta({\bm x})$ is essentially complex-valued if
the condensate contains vortices, e.g., $\zeta({\bm x}) \sim e^{i\kappa \theta}$
where $\theta$ is an angle around a vortex line and $\kappa$ is a winding number.
Equation (\ref{psi}) is substituted into Eq.~(\ref{S}), and the action is
rewritten in terms of $\zeta ({\bm x})$ and $\varphi(x)$ as follows:
\begin{equation}
S = S_0 + S_1+S_2 +S_{3,4} \, ,
\end{equation}
where
\label{Seps}
\begin{eqnarray}
S_0 &=&  \int \! d^4 x \left[ \zeta^\ast ({\bm x}) \{ - K - V
({\bm x}) + \mu \} \zeta ({\bm x}) - \frac{g}{2} | \zeta ({\bm x})|^4
\right] \, , \\
S_1 &=& \int \! d^4 x \, \Bigl\{
\zeta^\ast ({\bm x}) \left[ - K - V ({\bm x}) + \mu - g |\zeta ({\bm
x})|^2 \right] \varphi (x) \nn \\
&&+ \varphi^{\dag} (x) \left[ - K - V ({\bm
x})+ \mu - g |\zeta ({\bm x})|^2 \right] \zeta ({\bm x}) \Bigr\} \, , \\
S_2 &=&  \int \! d^4 x \, \Bigl\{ \varphi^\dag (x) \left[ T -
K - V ({\bm x}) + \mu \right] \varphi (x) \nn \\
&& - \frac{g}{2} \left[ 4 \,
|\zeta (\bm{x})|^2 \varphi^\dag (x) \varphi (x) + {\zeta^\ast}^2(\bm{x})
\varphi^2 (x) + \zeta^2 (\bm{x}) \varphi^{\dag 2} (x) \right] \Bigr\} \,
, \\
S_{3,4} &=&  \int \! d^4 x \, \Bigl\{ - g \left[ \zeta (\bm{x})
\varphi^{\dag 2} (x) \varphi (x) + \zeta^\ast (\bm{x}) \varphi^\dag (x)
\varphi^2 (x) \right] - \frac{g}{2} \varphi^{\dag 2} (x) \varphi^2 (x)
\Bigr\} \, . \label{Se} \nn \\
\end{eqnarray}

At the tree level, the c-number function $\zeta ({\bm x})$ satisfies
\begin{equation}
\left[K + V ({\bm x}) - \mu + g |\zeta ({\bm x})|^2 \right] \zeta ({\bm
 x}) = 0 \, , \label{eqv}
\end{equation}
which is called GP equation \cite{GP}.
The condensate particle number $N_{\rm c}$ is given by
\begin{equation}
N_{\rm c} = \int d^3 x \, |\zeta (\bm{x})|^2. \label{Nc}
\end{equation}

We move to the canonical formalism in the interaction representation.
The total Hamiltonian of the system is now written as
\begin{equation}
\hat{H} = \hat{H}_0 + \hat{H}_{\rm int} + \mathrm{const}. \, , \label{Heps}
\end{equation}
where
\begin{eqnarray}
\hat{H}_{0}
&=& \int \! d^3 x \, \Bigl\{ \hat{\varphi}^\dag (x) \{ K + V ({\bm x})
- \mu \} \hat{\varphi} (x) \nn \\ 
&&+ \frac{g}{2} \left[ 4 |\zeta ({\bm x})|^2
 \hat{\varphi}^\dag (x) \hat{\varphi} (x) +{\zeta^\ast}^2 ({\bm x})
 \hat{\varphi}^2 (x) +\zeta^2 ({\bm x}) \hat{\varphi}^{\dag 2} (x)
 \right] \Bigr\} \, ,  \label{eqn:h0}\\
\hat{H}_{\rm int} &=& \int \! d^3 x \, \left\{
g\left[ \zeta ({\bm x}) \hat{\varphi}^{\dag 2} (x) \hat{\varphi} (x) +
 \zeta^\ast ({\bm x}) \hat{\varphi}^\dag (x) \hat{\varphi}^2 (x) \right]
+ \frac{g}{2} \hat{\varphi}^{\dag 2} (x) \hat{\varphi}^2 (x) \right\} \,
 . \nn \\
\end{eqnarray}
It is emphasized that the CCRs,
\begin{equation}
 [ \hat{\varphi} ({\bm x},t), \hat{\varphi}^\dag ({\bm x'},t) ] =
\delta^3 (\bm{x} - \bm{x}') \label{CCR1}
\end{equation}
and
\begin{equation}
[\hat{\varphi} ({\bm x},t), \hat{\varphi} ({\bm x'},t) ] =
[ \hat{\varphi}^\dag ({\bm x},t), \hat{\varphi}^\dag ({\bm x'},t) ] =0
\, , \label{CCR2}
\end{equation}
must hold in consistent QFT.

We introduce the new function $f({\bm x})$ which is defined as
\begin{equation}
f ({\bm x}) = \frac{1}{\sqrt{N_{\mathrm{c}}}} \zeta ({\bm x}) \, ,
\end{equation}
which implies that $f(\bm{x})$ is normalized to be unity:
\begin{equation}
\int \! d^3 x \, |f(\bm{x})|^2 = 1 \, .
\end{equation}
We can rewrite the GP equation (\ref{eqv}) with $f({\bm x})$ as
\begin{equation}
{\mathcal L} ({\bm x}) f (\bm{x}) - {\mathcal M} ({\bm x}) f^* (\bm{x}) = 0 \, ,
\label{eqn:GPf}
\end{equation}
where the following notations are introduced:
\begin{eqnarray}
{\mathcal L} ({\bm x}) & =&  K + V ({\bm x}) - \mu + 2 g N_{\mathrm{c}}
|f (\bm{x})|^2 \, , \\
{\mathcal M} ({\bm x}) & = & g N_{\mathrm{c}} f^2 (\bm{x}) \, .
\end{eqnarray}

\section{Real eigenvalues of Bogoliubov-de Gennes Equations and
 Quasi-Particle Picture}\label{sec-BdG}

In this section, we review the BdG approach with real eigenvalues
including zero-mode \cite{Lewen,Matsu}.
We consider the following set of equations for excitation modes,
called the BdG equations:
\begin{eqnarray}
{\mathcal L} ({\bm x}) u_n (\bm{x}) - {\mathcal M} ({\bm x}) v_n (\bm{x})
& = & E_{n} u_{n} (\bm{x}) \, ,  \label{eqn:BdG1} \\
{\mathcal L} ({\bm x}) v_n (\bm{x}) - {\mathcal M}^* ({\bm x}) u_n (\bm{x})
 &= & - E_n v_n (\bm{x}) \label{eqn:BdG2}
\end{eqnarray}
for $n = 1, 2, \cdots$.  
For the zero-mode, while $f ({\bm x})$ satisfies Eq.~(\ref{eqn:GPf}),
another function denoted by $h ({\bm x})$ has to be introduced as
\begin{equation}
{\mathcal L} ({\bm x}) h ({\bm x}) + {\mathcal M} ({\bm x}) h^* ({\bm x}) =
 \frac{1}{I} f ({\bm x}) \, ,  \label{h}
\end{equation}
where $I$ is a real constant which will be determined by the condition
(\ref{fasth}). Here we assumed that the phase part of $h({\bm x})$ is
the same as that of $f ({\bm x})$. This does not affect the generality of our
discussion in this paper.
The eigenfunctions $u_n ({\bm x})$, $v_n ({\bm x})$, $f ({\bm x})$ and
$h({\bm x})$ satisfy the following orthogonal and completeness
conditions:
\begin{eqnarray}
 \int \! d^3 x \left[ u_n^\ast ({\bm x}) u_{n'} ({\bm x}) - v_n^\ast
 ({\bm x}) v_{n'} ({\bm x}) \right] & =& \delta_{nn'}
 \label{eqn:orthuv1} \, , \\
 \int \! d^3 x \left[ u_n ({\bm x}) v_{n'} ({\bm x}) - v_n ({\bm x})
 u_{n'} ({\bm x}) \right] & =& 0 \, , \label{eqn:orthuv2} \\
 \int \! d^3 x \, f^\ast ({\bm x}) h ({\bm x}) & = &\frac{1}{2}
 \, , \label{fasth} \\
\int \! d^3 x \, \left[
u_n^\ast({\bm x})f({\bm x}) - v_n^\ast({\bm x})f^\ast({\bm x})\right]&=&0 
\, ,\label{uvf}\\
\int \! d^3 x \, \left[
u_n^\ast({\bm x})h({\bm x}) + 
v_n^\ast({\bm x})h^\ast({\bm x})\right] &=&0\, ,\label{uvh}
\end{eqnarray}
and
\label{uvcomp}
\begin{eqnarray}
\sum_{n=1}^\infty \left[ u_n({\bm x})u_n^\ast({\bm x'})-
v_n^\ast({\bm x})v_n({\bm x'})\right] +
\left[ f({\bm x})h^*({\bm x'})+h({\bm x})f^*({\bm x'}) \right] &=&
 \delta^3 ({\bm x}-{\bm x'}) \, ,  \nn \\ \\
\sum_{n=1}^\infty \left[ u_n({\bm x})v_n^\ast({\bm x'})-
v_n^\ast({\bm x})u_n({\bm x'})\right]-
\left[  f({\bm x})h({\bm x'})-h({\bm x})f({\bm x'})\right] &=& 0 \, .
\end{eqnarray}

For real eigenvalues $E_n$, it is known that the quasi-particle picture
is obtained by diagonalization of the unperturbed Hamiltonian
(\ref{eqn:h0}).
The expansions of the quantum fields with respect to the annihilation- and
creation-operators of the quasi-particle $\hat{\alpha}_n$ and
$\hat{\alpha}_n^\dag$ are given as
\begin{eqnarray}
 \hat{\varphi} (x) & = \hat{\mathcal P} (t) h ({\bm x}) - i \hat{\mathcal Q}
  (t) f ({\bm x}) + \sum_{n = 1}^{\infty} \left[ u_n ({\bm x}) \hat{\alpha}_n (t) -
 v^\ast_n ({\bm x}) \hat{\alpha}_n^\dag (t) \right] \, ,
 \label{vphialpha} \\
 \hat{\varphi}^\dag (x) & = \hat{\mathcal P} (t) h^\ast ({\bm x}) + i
  \hat{\mathcal Q} (t) f^\ast ({\bm x}) + \sum_{n = 1}^{\infty} \left[ u_n^\ast ({\bm x})
 \hat{\alpha}_n^\dag (t) - v_n ({\bm x}) \hat{\alpha}_n (t) \right] \, .
 \label{vphidagalpha}
\end{eqnarray}
The operators $\hat{\alpha}_n (t)$ and $\hat{\alpha}_n^\dag (t)$
are subject to the canonical commutation relations: $[
\hat{\alpha}_n (t), \hat{\alpha}_{n'}^\dag (t)] = \delta_{nn'}$ and $[
\hat{\alpha}_n (t), \hat{\alpha}_{n'} (t) ] = [ \hat{\alpha}_n^\dag (t),
\hat{\alpha}_{n'}^\dag (t) ] = 0$, and $\hat{\mathcal Q} (t)$ and $\hat{\mathcal
P} (t)$, which are called ``quantum coordinates'' (QCs), satisfy the
canonical commutation relations: $[\hat{\mathcal Q} (t), \hat{\mathcal P} (t)] =
i$ and other commutations vanish.
The relation between QCs and zero-energy particle
mode is studied in Ref.~\cite{Mine}.

Then, one can find the unperturbed Hamiltonian (\ref{eqn:h0}) is
diagonalized in terms of $\hat{\alpha}_n^\dag (t)$, $\hat{\alpha}_n
(t)$, $\hat{\mathcal Q} (t)$ and $\hat{\mathcal P} (t)$ as
\begin{equation}
\hat{H}_0 = \frac{1}{2I} \hat{\mathcal P}^2 +  \sum_{n=1}^{\infty}
E_n \hat{\alpha}_n^\dag \hat{\alpha}_n + {\rm const.}
\end{equation}
This determines the time dependence of the operators:
\begin{eqnarray}
 \hat{\alpha}_n (t) & = & \hat{\alpha}_n e^{- i E_n t} \, , \quad
 \hat{\alpha}_n^\dag (t) = \hat{\alpha}_n^\dag e^{i E_n t} \, , \nn \\
 \hat{\mathcal Q} (t) & = & \hat{\mathcal Q} + \hat{\mathcal P} t \, , \quad
 \hat{\mathcal P} (t) = \hat{\mathcal P} \, .
\end{eqnarray}

\section{Doublet Notation}\label{sec-doublet}

We introduce the doublet notation,
\begin{equation}
 r ({\bm x}) =
\begin{pmatrix}
r_1(\bm{x}) \\
r_2(\bm{x})
\end{pmatrix}
\, .
\end{equation}
Using this doublet notation, we define the following ``inner product''
for a pair of any doublets as
\begin{eqnarray}
(r, s) & \equiv & \int \! d^3 x \, r^\dag ({\bm x}) \sigma_3 s
 ({\bm x}) \nn \\
& =& \int \! d^3 x \,
\begin{array}{cc}
\left( r^*_1(\bm{x}) \, r^*_2(\bm{x}) \right)
\end{array}
\left(
\begin{array}{cc}
1 & 0 \\
0 & -1
\end{array}
\right)
\left(
\begin{array}{c}
s_1(\bm{x}) \\
s_2(\bm{x})
\end{array}
\right)
\nn \\
& = & \int \! d^3 x \,
[ r^*_1(\bm{x}) \, s_1(\bm{x}) - r^*_2(\bm{x}) \, s_2(\bm{x}) ] \, ,
\end{eqnarray}
where $\sigma_i$ represents the $i$-th Pauli matrix.

We also define a (squared) ``norm'' of $r$ as
\begin{equation}
\| r \|^2
\equiv
(r , \, r)
= \int \! d^3 x \left[ |r_1 ({\bm x})|^2 - |r_2 ({\bm x})|^2
\right] \, .
\end{equation}
We note that this ``norm'' can be zero or negative due to
the metric $\sigma_3$.

The doublets of eigenfunctions of the BdG equations are defined as
follows:
\begin{eqnarray}
 x_n ({\bm x}) & \equiv &
\left(
\begin{array}{c}
   u_n ({\bm x}) \\
   v_n ({\bm x})
\end{array}
\right)
 \qquad (n = 1, 2, \cdots) \, , \\
 x_0 ({\bm x}) & \equiv &
\left(
\begin{array}{c}
   f ({\bm x}) \\
   f^\ast ({\bm x})
\end{array}  
\right) \, ,
 \\
x_{-1} ({\bm x}) & \equiv &
\left(
\begin{array}{c}
   h ({\bm x}) \\
   - h^\ast ({\bm x})
\end{array} 
\right)\, .
\end{eqnarray}
We also introduce the symbol of 
\begin{equation}
y_n ({\bm x})  \equiv \sigma_1 x_n^\ast ({\bm x}) =
\begin{pmatrix}
  v_n^\ast ({\bm x}) \\
  u_n^\ast ({\bm x})
\end{pmatrix}
\qquad (n = 1, 2, \cdots) \, . 
\end{equation}

Under the doublet notation, Eqs.~(\ref{eqn:BdG1}) and (\ref{eqn:BdG2}) read
as
\begin{equation}
T ({\bm x}) x_n ({\bm x}) = E_n x_n ({\bm x}) \, ,  \label{eqn:dBdG}
\end{equation}
where
\begin{equation}
T ({\bm x}) =
\left(
\begin{array}{cc}
{\mathcal L} ({\bm x}) & -{\mathcal M} ({\bm x}) \\
{\mathcal M}^* ({\bm x}) & -{\mathcal L} ({\bm x})
\end{array}
\right)
\, .
\end{equation}
Using this representation we can rewrite the GP equation (\ref{eqn:GPf})
as 
\begin{equation}
T ({\bm x}) \, x_0 ({\bm x}) = 0 \, ,
\end{equation}
and Eq.~(\ref{h}) as
\begin{equation}
T ({\bm x}) x_{-1} ({\bm x}) = \frac{1}{I}x_0 ({\bm x}) \, . \label{eqn:hh}
\end{equation}
Note that $\|x_0\|^2=\|x_{-1}\|^2=0$. 
The orthogonal conditions (\ref{eqn:orthuv1})--(\ref{uvh})
and the completeness conditions are rewritten, respectively, as
\begin{eqnarray}
 (x_n, x_{n'}) & = &\delta_{nn'} \, , \\
 (y_n , y_{n'}) & =& - \delta_{nn'} \, , \\
 (y_n , x_{n'}) & =& 0 \, , \\
 (x_0 , x_{-1}) & =& 1 \, , \\
 (x_n , x_0) & =& (x_n , x_{-1})=0 \, ,
\end{eqnarray}
for $n,n' = 1, 2, \cdots$, and
\begin{eqnarray}
 \sum_{n=1}^{\infty} \left[ x_n ({\bm x}) \, x_n^\dag ({\bm x'}) - y_n
 ({\bm x}) \, y_n^\dag ({\bm x'}) \right] + x_0 ({\bm x}) x_{-1}^\dag ({\bm x'}) &+& x_{-1} ({\bm x}) x_0^\dag
({\bm x'}) \nn \\
&&= \sigma_3 \delta^3 ({\bm x} - {\bm x'}) \, . 
 \label{eqn:comp}
\end{eqnarray}
The CCRs (\ref{CCR1}) and (\ref{CCR2}) are rewritten in
the doublet expression as
\begin{equation}
[\hat{\Phi}(\bm{x},t), \hat{\Phi}^{\dag}(\bm{x}',t)] =
\sigma_3 \delta^3(\bm{x} - \bm{x}') \, ,
\end{equation}
where
\begin{equation}
 \hat{\Phi} (x) =
\left(
\begin{array}{c}
\hat{\varphi}(x) \\
\hat{\varphi}^{\dag}(x)
\end{array}
\right)
\, . \label{doubletvphi}
\end{equation}

\section{Complex Eigenvalues of Bogoliubov-de Gennes Equations}\label{sec-complexBdG}

Hereafter, we consider the situation in which the BdG equations have complex
eigenvalues. Many numerical calculations \cite{Pu,Garay,Skryabin,Mettenen,Kawaguchi}
show that some complex eigenvalues appear in the presence of highly quantized
vortices. We show some basic mathematical properties of the eigenfunctions, and
then a symmetric property for the BdG equations.
Finally we give a complete set including the eigenfunctions which belong to the
complex eigenvalues.
This plays a crucial role in constructing QFT involving the complex eigenvalues.

\subsection{Mathematical Preliminaries}\label{sec-CE}

First, it is important to see that 
\textit{if $E_n$ is a complex number {\rm(${\rm Im}(E_n) \neq 0$)}, then, $ \| x_n \|^2 =0 $.}
This fact is shown from the following relation: 
\begin{equation}
E^\ast_n (x_n , \, x_n) = (Tx_n, \, x_n) = (x_n, \, Tx_n) = E_n (x_n ,
 \, x_n) \, .
\end{equation}

The same discussion of this fact is given in Ref.~\cite{Garay}. The
similar analyses, related to this fact
as well as the next statement shown below,
are found for the special cases of the BECs with single
multiply quantized vortex \cite{Skryabin,Kawaguchi}. 


Next, let us see that \textit{if}
\be
(x_k , \, x_m)\neq 0 \, , \label{eqn:nonortho}
\ee
\textit{then, $E^*_k = E_m$.} Here $x_k$ and $x_m$ are the eigenvectors whose 
eigenvalues are $E_k$ and $E_m$, respectively. This is shown from
\begin{equation}
E^*_k (x_k , \, x_m) = (T x_k, \, x_m) = (x_k, \, T x_m) = E_m (x_k , \,
 x_n) \, .
\end{equation}
Note that $x_k$ and $x_m$ are not necessarily complex modes. 

We recall that the condensate can have complex eigenvalues
when it has flow in a optical lattice or a highly quantized vortex.
In those cases, the condition (\ref{eqn:nonortho}) 
holds \cite{Wu,Skryabin,Kawaguchi}, 
and the complex eigenvalues arise as a conjugate pair.

As a contraposition of the above statement, we can see that
\textit{if}
\begin{equation}
E^\ast_k \neq E_m \label{eqn:condi} \, ,
\end{equation}
\textit{then, $(x_k, \, x_m) = 0$.}
This provides the sufficient condition for the orthogonality
between different modes.
In particular, two complex modes are orthogonal to each other 
if they satisfy the condition (\ref{eqn:condi}). 
We can also see that any complex mode is orthogonal to all
real modes. 




\subsection{Symmetry for Bogoliubov-de Gennes
  Equations}\label{sec-sym} 
It is known that there exists a symmetry for the BdG equations when their
eigenvalues are real \cite{Matsu}. For the BdG equations with
complex eigenvalues, we find that
\textit{if $x_n$ is an eigenvector which belongs to eigenvalue $E_n$:}
\begin{equation}
T ({\bm x}) x_n ({\bm x}) =E_n  x_n ({\bm x}) \, ,
\end{equation}
\textit{then, the following relation holds,}
\begin{equation}
T ({\bm x}) y_n ({\bm x}) = -E^*_n  y_n ({\bm x}) \, .
\end{equation}
This implies that $y_n$ is also an eigenvector, belonging to the eigenvalue
 $-E^*_n$. Note that the following relation also holds,
\begin{equation}
\|y_n \|^2 = - \|x_n \|^2.
\end{equation}
To check the above relation, one uses $\sigma_1 T^\ast \sigma_1 = -T$.

The similar analyses are found for the special cases
 of the BEC system with a single multiply quantized 
vortex \cite{Skryabin,Kawaguchi}
and of the BEC system under the periodic potential \cite{Wu}.

\subsection{Complete Set with Complex Modes}
For QFT, we need a complete set in order to expand quantum fields.
In this subsection, we present a complete set with complex modes.
The Roman indices ($n$, $m$, ...) and the Greek ones ($\mu$, $\nu$, ...),
which are integers, 
are used for the real modes and for the complex ones, respectively,
in our discussions below. 

Let us assume that there is a set of some pairs of complex modes, 
say $\{x_\mu , x_{\ast \mu}\}$, which satisfy
\ba
T x_\mu &=& E_\mu x_\mu \, , \\
T x_{\ast \mu} &=& E_{\ast \mu} x_{\ast \mu} \, .
\ea
For different $\mu$,
it is also assumed that $E_\mu$ is equal to neither $E_\nu $ nor $E_{\ast \nu}$
for $\mu \neq\nu$, meaning no degeneracy in complex modes. We adjust the normalization
constants, so that 
the ``inner product'' is simply given by 
\begin{equation}
(x_\mu \, , \, x_{\ast \nu}) = \delta_{\mu \nu}. \label{eqn:xk_norm1}
\end{equation}
The results of Subsec.~\ref{sec-CE} imply 
$E_{\mu}^\ast = E_{\ast \mu}$ and 
\begin{equation}
(y_\mu \, , \, y_{\ast \nu}) = - \delta_{\mu \nu} \, ,
 \label{eqn:yk_norm1} 
\end{equation}
where
\begin{eqnarray}
 y_\mu & \equiv \sigma_1 x^\ast_\mu \, , \\
 y_{\ast \mu} & \equiv \sigma_1 x_{\ast \mu}^\ast \, .
\end{eqnarray}
For later convenience, we introduce the elements of $x_\mu$ and
$x_{\ast \mu}$ as 
\begin{eqnarray}
 x_\mu & =
\left(
\begin{array}{c}
 u_\mu ({\bm x}) \\
 v_\mu ({\bm x})
\end{array} 
\right)\, , 
\\
 x_{\ast \mu} & =
\left(
\begin{array}{c}
 u_{\ast \mu} ({\bm x}) \\
 v_{\ast \mu} ({\bm x})
\end{array} 
\right)
\, . 
\end{eqnarray}

This way we can construct the complete set
which includes the complex modes and is consistent with all the
mathematical properties of the complex modes presented in Section
\ref{sec-CE} and Eqs.~(\ref{eqn:xk_norm1}) and (\ref{eqn:yk_norm1})
as
\begin{eqnarray}
\lefteqn{\quad \sum_n
\left[
x_n (\bm{x}) x_n^{\dag}(\bm{x}')
-
y_n (\bm{x}) y_n^{\dag}(\bm{x}')
\right]+
x_0(\bm{x}) x^{\dag}_{-1}(\bm{x'})
+
x_{-1}(\bm{x}) x^{\dag}_0(\bm{x'})
}  &&
 \nn \\
& \quad {}& +
\sum_\mu \left[
x_\mu (\bm{x}) x^\dag_{\ast \mu} (\bm{x'})
+
x_{\ast \mu} (\bm{x}) x^\dag_\mu (\bm{x'})
-
y_\mu (\bm{x}) y^\dag_{\ast \mu} (\bm{x'})
-
y_{\ast \mu} (\bm{x}) y^\dag_\mu (\bm{x'})
\right] \nn \\
& & \quad =
\sigma_3 \delta^3 (\bm{x} - \bm{x}') \, . \label{eqn:complete}
\end{eqnarray}


\section{Representation of Free Hamiltonian and Expansion of Quantum
 Fields in Terms of Complex Mode Wave Functions}\label{sec-Hamiltonian}

\subsection{Representation of Free Hamiltonian and its Properties}
As we have the complete set with complex modes, we can expand the field
operators. 
We discuss first in the Schr{\" o}dinger picture and next move to the
interaction picture. 

According to the completeness condition (\ref{eqn:complete}), 
the field operators are expanded in the Schr{\" o}dinger picture as
follows:
\begin{eqnarray}
 \hat{\varphi}({\bm x})
&=&
\sum_n
\left[
\hat{\alpha}_n u_n(\bmx)
-
\hat{\alpha}^{\dag}_n v^{\ast}_n(\bmx)
\right]
+
\hat{\mathcal P} h(\bmx) - i \hat{\mathcal Q} f(\bmx) \nn \\
&&{} +
\sum_\mu \left[
\hat{A}_\mu u_\mu (\bm{x})
+
\hat{B}_\mu u_{\ast \mu} (\bm{x})
-
\hat{A}^\dag_\mu v^\ast_\mu (\bm{x})
-
\hat{B}_\mu^\dag v_{\ast \mu}^\ast (\bm{x}) \right] \, , \\
\hat{\varphi}^{\dag}({\bm x}) 
&=&
\sum_n
\left[
\hat{\alpha}^{\dag}_n u^{\ast}_n(\bmx)
-
\hat{\alpha}_n v_n(\bmx)
\right]
+
\hat{\mathcal P} h^{\ast}(\bmx) + i \hat{\mathcal Q} f^{\ast}(\bmx) \nn \\
&&{}+ \sum_\mu \left[
\hat{A}^\dag_\mu u^\ast_\mu (\bm{x})
+
\hat{B}^\dag_\mu u^\ast_{\ast \mu} (\bm{x})
-
\hat{A}_\mu v_\mu (\bm{x})
-
\hat{B}_\mu v_{\ast \mu} (\bm{x}) \right] \, , 
\end{eqnarray}
where the sets of operators $\{{\hat \alpha}_n, {\hat \alpha}^\dag_n\}$
and $\{ \hat{\mathcal Q},\hat{\mathcal P}\}$ are associated with the real modes
and the zero-mode as discussed in Sec.~\ref{sec-BdG}, respectively,
while the operators $\hat{A}_\mu$, $\hat{A}^\dag_\mu$, $\hat{B}_\mu$, and
$\hat{B}_\mu^\dag$ are newly introduced in connection with the complex modes. 
To be consistent with the CCRs, the new operators for complex modes
have to satisfy the following commutation 
relations \cite{Garay,RK}: 
\begin{eqnarray}
\left[ \hat{A}_\mu, \hat{B}^\dag_\nu \right] &=& \delta_{\mu \nu} \, , \label{ABdag}
 \\ 
\left[ \hat{A}_\mu, \hat{A}^\dag_\nu \right] &= &0 \, , \\
\left[ \hat{B}_\mu, \hat{B}^\dag_\nu \right] &= &0 \, , \\
\left[ \hat{A}_\mu, \hat{B}_\nu \right] &= &0 \, . \label{AB}
\end{eqnarray}

Using the above representations, one can rewrite the free Hamiltonian
$\hat{H}_0$ as
\begin{eqnarray}
\hat{H}_0
&=
\frac{1}{2I} \hat{\mathcal P}^2 +
\sum_n E_n \hat{\alpha}_n^{\dag} \hat{\alpha}_n
 + \sum_\mu \left[
E^\ast_\mu \hat{A}^\dag_\mu \hat{B}_\mu + E_\mu \hat{B}^\dag_\mu
 \hat{A}_\mu \right] \, . 
\end{eqnarray}
One can easily check the hermiticity of $\hat{H}_{0}$ in this representation.
For later convenience, we denote the complex mode sector of $\hat{H}_0$
by $\hat{H}_0^{({\rm c})}$:
\begin{equation}
 \hat{H}_0^{({\rm c})} = \sum_\mu \left[ E^\ast_\mu \hat{A}^\dag_\mu
 \hat{B}_\mu + E_\mu \hat{B}^\dag_\mu \hat{A}_\mu \right] \, ,
\end{equation}
and we call $\hat{H}_0^{({\rm c})}$ \textit{the complex mode sector
Hamiltonian}.
One finds the time evolution of $\hat{A}_\mu$ and $\hat{B}_\mu$ as follows:
\begin{eqnarray}
 A_\mu (t) & = \rme^{- \rmi E_\mu t} A_\mu, \label{AteA} \\
 A^\dag_\mu (t) & = \rme^{\rmi E^\ast_\mu t} A^\dag_\mu , \\
 B_\mu (t) & = \rme^{- \rmi E^\ast_\mu t} B_\mu, \\
 B^\dag_\mu (t) & = \rme^{\rmi E_\mu t} B^\dag_\mu. \label{BdteBd}
\end{eqnarray}
Note that this representation is consistent with the fact that
$\hat{H}_0$ is time-independent. 
Thus the emergence of the complex eigenvalues
does not imply the instability of the system directly. 
This point will be discussed in detail in Sec.~\ref{sec-Phys}. 

Now, we obtain the quantum field in the interaction representation,
\begin{eqnarray}
 \hat{\varphi}(x)
&=&
\sum_n
\left[
\hat{\alpha}_n e^{- i E_n t} u_n(\bmx)
-
\hat{\alpha}^{\dag}_n e^{i E_n t} v^{\ast}_n(\bmx)
\right]
+
\hat{\mathcal P} h(\bmx) - i (\hat{\mathcal Q} + \hat{\mathcal P} t ) f(\bmx) \nn \\
&& {}
+ \sum_\mu \Big[
\hat{A}_\mu e^{- i E_\mu t} u_\mu (\bm{x})
+
\hat{B}_\mu e^{- i E^\ast_\mu t} u_{\ast \mu} (\bm{x}) \nn \\
&& \qquad \quad -
\hat{A}^\dag_\mu e^{i E^\ast_\mu t} v^\ast_\mu (\bm{x})
-
\hat{B}^\dag_\mu e^{i E_\mu t} v_{\ast \mu}^{\ast}(\bm{x}) \Big]
 \label{intvphi} \\ 
\hat{\varphi}^{\dag}(x)
&=&
\sum_n
\left[
\hat{\alpha}^{\dag}_n e^{i E_n t} u^{\ast}_n(\bmx)
-
\hat{\alpha}_n e^{- i E_n t} v_n(\bmx)
\right]
+
\hat{\mathcal P} h^{\ast}(\bmx) + i (\hat{\mathcal Q} + \hat{\mathcal P} t )
 f^{\ast}(\bmx) \nn \\
&& {}
+ \sum_\mu \Big[
\hat{A}^\dag_\mu e^{i E^\ast_\mu t} u^\ast_\mu (\bm{x})
+
\hat{B}^\dag_\mu e^{i E_\mu t} u^\ast_{\ast \mu} (\bm{x}) \nn \\
&& \qquad \quad -
\hat{A}_\mu e^{- i E_\mu t} v_\mu (\bm{x})
-
\hat{B}_\mu e^{- i E^\ast_\mu t} v_{\ast \mu} (\bm{x}) \Big] \, 
 . \label{intvphidag} 
\end{eqnarray}
One can easily check that they satisfy the CCRs,
\begin{eqnarray}
 \left[\hat{\varphi} (\bmx, t), \hat{\varphi}^\dag (\bmx', t)\right] & = & \delta^3
 (\bmx - \bmx') \, , \\
 \left[\hat{\varphi} (\bmx, t), \hat{\varphi} (\bmx', t)\right] & = &
 \left[\hat{\varphi}^\dag (\bmx, t), \hat{\varphi}^\dag (\bmx', t)\right] = 0 \, .
\end{eqnarray}


In order to study the complex mode operators $\hat{A}_\mu$ and
$\hat{B}_\mu$ further, we represent them by the two kinds of bosonic
operators \cite{Garay}. 
We introduce the operators $b_\mu$, $b^\dag_\mu$, $\tilde{b}_\mu$ and 
$\tilde{b}^\dag_\mu$ as follows:
\begin{eqnarray}
\hat{A}_\mu &= \frac{1}{\sqrt{2}} (\hat{b}_\mu + i \tilde{b}^\dag_\mu)
 \, , \label{Abbtild} \\
\hat{B}_\mu &= \frac{1}{\sqrt{2}} (\hat{b}_\mu - i \tilde{b}^\dag_\mu)
 \, , \label{Bbbtild}
\end{eqnarray}
or equivalently,
\begin{eqnarray}
\hat{b}_\mu &= \frac{1}{\sqrt{2}} (\hat{A}_\mu + \hat{B}_\mu) \, , \\ 
\tilde{b}_\mu &= \frac{i}{\sqrt{2}} (\hat{A}^\dag_\mu -
 \hat{B}^\dag_\mu) \, . 
\end{eqnarray}
The above definitions bring 
the following commutation relations:
\begin{eqnarray}
\left[ \hat{b}_\mu \, , \, \hat{b}^\dag_\nu \right] & = \delta_{\mu \nu} \, , \\ 
\left[ \tilde{b}_\mu \, , \, \tilde{b}^\dag_\nu \right] &= \delta_{\mu \nu} \, ,
\end{eqnarray}
with the other vanishing commutations.
Similarly, one can easily check that the bosonic representations
(\ref{Abbtild}) and (\ref{Bbbtild}) satisfy the relations
(\ref{ABdag})--(\ref{AB}).

Now, we can rewrite the Hamiltonian $\hat{H}_0^{({\rm c})}$ in terms of
$\hat{b}_\mu$ and $\tilde{b}_\mu$ as follow \cite{Kawaguchi}:
\begin{eqnarray}
& \hat{H}_0^{({\rm c})} = \sum_\mu \left[
\mathrm{Re}(E_\mu) (\hat{b}^\dag_\mu \hat{b}_\mu - \tilde{b}^\dag_\mu 
\tilde{b}_\mu) 
+\mathrm{Im}(E_\mu) (\hat{b}^\dag_\mu \tilde{b}^\dag_\mu + \tilde{b}_\mu 
\hat{b}_\mu) \right] \, . \label{eqn:H_b}
\end{eqnarray}

Let us see that $\hat{H}_0^{({\rm c})}$ is not diagonalizable
using usual bosonic representation.
Now we consider the Bogoliubov transformations, given by
\begin{eqnarray}
\hat{b}_\mu &= c_\mu \hat{\xi}_\mu - s_\mu \tilde{\xi}^\dag_\mu \, , \\ 
\tilde{b}_\mu &= c_\mu \tilde{\xi}_\mu - s_\mu \hat{\xi}^\dag_\mu \, ,
\end{eqnarray}
where the real numbers $c_\mu$ and $s_\mu$ satisfy
\begin{equation}
c_\mu^2 - s_\mu^2 = 1 \, , \label{eqn:cs1}
\end{equation}
which ensures the bosonic commutation relations for
 $\hat{\xi}_\mu$ and $\tilde{\xi}_\mu$,
\begin{eqnarray}
\left[ \hat{\xi}_\mu \, , \, \hat{\xi}^\dag_\nu \right] & = \delta_{\mu \nu} \, ,
 \\ 
\left[ \tilde{\xi}_\mu \, , \, \tilde{\xi}^\dag_\nu \right] & = \delta_{\mu \nu} \,
 , 
\end{eqnarray}
and zero for the other commutations.
Then $\hat{H}_0^{(\mathrm{c})}$ is rewritten in terms of $\hat{\xi}$ and
 $\tilde{\xi}$ as
\begin{eqnarray}
\hat{H}_0^{(\mathrm{c})} &= \sum_\mu \Bigl[ 
\left\{ \mathrm{Re}(E_\mu) - 2 c_\mu s_\mu \mathrm{Im}(E_\mu) \right\}
 \hat{\xi}^\dag_\mu \, \hat{\xi}_\mu -
\left\{
\mathrm{Re}(E_\mu) + 2 c_\mu s_\mu \mathrm{Im}(E_\mu) \right\}
 \tilde{\xi}^\dag_\mu \, \tilde{\xi}_\mu \nn \\
& \quad {} +
\mathrm{Im}(E_\mu) (c_\mu^2 + s_\mu^2) \hat{\xi}^\dag_\mu
 \tilde{\xi}^\dag_\mu +
\mathrm{Im}(E_\mu) (c_\mu^2 + s_\mu^2) \hat{\xi}_\mu \tilde{\xi}_\mu
 \Bigr] \, . 
\end{eqnarray}
The Hamiltonian $\hat{H}_0^{(\mathrm{c})}$ could be diagonal only if
 the following relation for each $\mu$ were fulfilled,
\begin{equation}
\mathrm{Im}(E_\mu) (c_\mu^2 + s_\mu^2) = 0 \, . 
\label{eqn:condition0}
\end{equation}
But from Eq.~(\ref{eqn:cs1}), we have $c_\mu^2 + s_\mu^2 = 2 s_\mu^2 + 1
 \ge 1$, and Eq.~(\ref{eqn:condition0}) would lead to $\mathrm{Im}(E_\mu) = 0$
for each $\mu$. 
Thus when the eigenvalue $E_\mu$ is complex, the Bogoliubov
 transformation can not diagonalize $\hat{H}_0^{(\mathrm{c})}$.
We therefore conclude that the complex mode sector Hamiltonian
$\hat{H}_0^{(\mathrm{c})}$ is not diagonalizable in the bosonic representation.

\subsection{Eigenstates of Complex Mode Sector Hamiltonian
  $\hat{H}_0^{(\mathrm c)}$}\label{sec-eigenstate}
In this subsection, we construct eigenstates of $\hat{H}_0^{({\mathrm c})}$,
starting from the vacuum of $\hat{b}_\mu$ and $\tilde{b}_\mu$.

First, we consider a transformation from the set of the bosonic
operators $\{\hat{b}_\mu, \hat{b}^\dag_\mu, \tilde{b}_\mu,
\tilde{b}^\dag_\mu\}$ to the set of the complex mode operators
$\{\hat{A}_\mu, \hat{A}^\dag_\mu, \hat{B}_\mu, \hat{B}^\dag_\mu \}$.
One finds the following explicit transformations:
\begin{eqnarray}
\hat{A}_\mu &= \hat{W}_\mu \hat{b}_\mu \hat{W}^{-1}_\mu = i \hat{W}^{-1}_\mu \tilde{b}^\dag_\mu \hat{W}_\mu \, ,
 \label{AUtildbU} \\ 
\hat{B}_\mu &= \hat{W}^{-1}_\mu \hat{b}_\mu \hat{W}_\mu  = - i \hat{W}_\mu \tilde{b}^\dag_\mu \hat{W}^{-1}_\mu \, ,
 \label{BUtildbU} 
\end{eqnarray}
and
\begin{eqnarray}
 \hat{A}^\dag_\mu & = - i \hat{W}_\mu \tilde{b}_\mu \hat{W}^{-1}_\mu
 = \hat{W}^{-1}_\mu \hat{b}^\dag_\mu \hat{W}_\mu \, , \label{AdagUbU} \\
 \hat{B}^\dag_\mu & = i \hat{W}^{-1}_\mu \tilde{b}_\mu \hat{W}_\mu
  = \hat{W}_\mu \hat{b}^\dag_\mu \hat{W}^{-1}_\mu \, , \label{BdagUbU}
\end{eqnarray}
where
\begin{equation}
 \hat{W}_\mu = \exp
\left[
i \frac{\pi}{4}
(
\hat{b}_\mu \tilde{b}_\mu - \hat{b}^\dag_\mu \tilde{b}^\dag_\mu
)
\right] \, .
\end{equation}
Note that the operator $W_\mu$ is not unitary but hermitian,
\begin{equation}
 \hat{W}^\dag_\mu = \hat{W}_\mu \, , \label{HermiteU}
\end{equation}
and its inverse operator is given as
\begin{equation}
 \hat{W}^{-1}_\mu = \exp
\left[
- i \frac{\pi}{4}
(
\hat{b}_\mu \tilde{b}_\mu - \hat{b}^\dag_\mu \tilde{b}^\dag_\mu
)
\right] \, .
\end{equation}
For later convenience, we introduce the notation of
\begin{eqnarray}
 \hat{W} & \equiv \bigotimes_\mu \hat{W}_\mu \, , \\
 \hat{W}^{-1} & \equiv \bigotimes_\mu \hat{W}_\mu^{-1} \, . 
\end{eqnarray}

Next, we construct states which are annihilated by $\hat{A}_\mu$,
$\hat{A}^\dag_\mu$, $\hat{B}_\mu$ and $\hat{B}^\dag_\mu$.
In order to find those states, we introduce a state
$\KKV$ defined by
\be
\KKV \equiv \bigotimes_\mu \KKV_\mu, 
\ee
where $\KKV_\mu$ is
the tensor product of the vacuum of $\hat{b}_\mu$ and $\tilde{b}_\mu$:
\begin{equation}
\KKV_\mu = \KV {}_{b\mu} \otimes \KV {}_{\tilde{b} \mu} \, ,
\end{equation}
where 
\begin{eqnarray}
 \left\{
\begin{array}{cc}
 \hat{b}_\mu \KV {}_{b \mu} & = 0 \, ,  \\
 \tilde{b}_\mu \KV {}_{\tilde{b} \mu} & = 0 \, .
\end{array}
 \right.
\end{eqnarray}
Obviously, the state $\KKV_\mu$ is annihilated by both $\hat{b}_\mu$ and 
$\tilde{b}_\mu$, i.e.,
\begin{equation}
 \left\{
\begin{array}{cc}
 \hat{b}_\mu \KKV_\mu & = 0 \, ,  \\
 \tilde{b}_\mu \KKV_\mu & = 0 \, .
\end{array}
 \right.
\end{equation}

Now, we introduce the states $\KV_A$ and $\KV_B$ defined by
\begin{eqnarray}
\KV_A &= \hat{W} \KKV \, ,
 \label{eqn:zeroA} \\ 
\KV_B &= \hat{W}^{-1} \KKV \, ,
 \label{eqn:zeroB} 
\end{eqnarray}
Because of hermiticity of $\hat{W}_\mu$ (\ref{HermiteU}), we obtain 
${}_A \BV$ and ${}_B \BV$ as 
\begin{eqnarray}
 {}_A \BV & = \BBV \hat{W} \, ,
 \\ 
 {}_B \BV & = \BBV \hat{W}^{-1}
 \, . 
\end{eqnarray}
One can easily find the following relations by using
Eqs.~(\ref{AUtildbU})--(\ref{BdagUbU}): 
\begin{eqnarray}
\hat{A}_\mu \KV {}_A &= 0 \, , \label{eqn:A0} \\
\hat{A}^\dag_\mu \KV {}_A &= 0 \, , \label{eqn:Adag0} \\
\hat{B}_\mu \KV {}_B &= 0 \, ,  \label{eqn:B0} \\
\hat{B}^\dag_\mu \KV {}_B &= 0 \, ,  \label{eqn:Bdag0}
\end{eqnarray}
and
\begin{eqnarray}
{}_A \BV \hat{A}_\mu &= 0 \, , \label{eqn:0A} \\
{}_A \BV \hat{A}^\dag_\mu &= 0 \, , \label{eqn:0dagA} \\
{}_B \BV \hat{B}_\mu &= 0 \, . \label{eqn:0B} \\
{}_B \BV \hat{B}^\dag_\mu &= 0 \, , \label{eqn:0dagB}
\end{eqnarray}
which are valid for arbitrary $\mu$. 
Hereafter we call $\KV_A$ and $\KV_B$ \textit{zero states} of $\hat{A}_\mu$
and $\hat{B}_\mu$, respectively.

These zero states are eigenstates of
$\hat{H}_0^{({\mathrm c})}$,
\begin{eqnarray}
\hat{H}_0^{({\mathrm c})} \KV {}_A & = - \left[ \sum_\mu E_\mu^{\ast}
 \right] \KV {}_A \, , \label{eqn:H0A} \\ 
\hat{H}_0^{({\mathrm c})} \KV {}_B & = - \left[ \sum_\mu E_\mu \right]
 \KV {}_B \, . \label{eqn:H0B}
\end{eqnarray}
Note the following relations for the bra states:
\begin{eqnarray}
{}_A \BV \hat{H}_0^{({\mathrm c})} & = - \left[ \sum_\mu E_\mu \right]
 {}_A \BV \, , \label{eqn:AH0} \\
{}_B \BV \hat{H}_0^{({\mathrm c})} & = - \left[ \sum_\mu E^\ast_\mu
 \right] {}_B \BV \, . \label{eqn:BH0}
\end{eqnarray}

Using Eqs.~(\ref{eqn:zeroA}) and (\ref{eqn:zeroB}),
one easily derives
\begin{equation}
{}_A \braket{0}{0} {}_B = {}_B \braket{0}{0} {}_A = 1 \, .
 \label{eqn:ABnorm} 
\end{equation}

It is important to notice that both of ${}_A \braket{0}{0} {}_A $
and ${}_B \braket{0}{0} {}_B $ diverge.
To show this fact, we rewrite ${}_A \braket{0}{0} {}_A $ in the bosonic
representation: 
\begin{eqnarray}
{}_A\braket{0}{0} {}_A & = & \prod_\mu \left[ {}_\mu \BBV
 \hat{W}^2_\mu \KKV {}_\mu \right] \nn \\
& = &
\prod_\mu \left[ {}_\mu \BBV e^{i \frac{\pi}{2} \left( \hat{b}_\mu
 \tilde{b}_\mu - \hat{b}^\dag_\mu \tilde{b}^\dag_\mu \right)} \KKV {}_\mu 
\right] \, . \label{eqn:0Anorm}
\end{eqnarray}
Define a new state $\ket{\theta}$ by
\begin{align}
\ket{\theta} 
& \equiv \bigotimes_\mu \left[ e^{- \theta \left( \hat{b}_\mu
 \tilde{b}_\mu - \hat{b}^\dag_\mu \tilde{b}^\dag_\mu \right)} \KKV
 {}_\mu \right] \nn \\ 
& = \bigotimes_\mu \left[ \exp (-\ln \cosh \theta) \exp \left(
 \hat{b}^\dag_\mu \tilde{b}^\dag_\mu \tanh \theta \right) \KKV {}_\mu
 \right] \, , \label{eqn:w2kv} 
\end{align}
where the constant term on the right-hand side 
is determined by the normalizatoin condition $\braket{\theta}{\theta} =
1$. 
One finds the limiting behavior
\begin{equation}
 \BBV \theta \rangle \rightarrow \infty \quad \left( \theta \rightarrow - i
  \frac{\pi}{2} \right) \, , 
\end{equation}
which implies that
${}_A \braket{0}{0} {}_A$ is divergent. 
Similarly, it can be proven that ${}_B \braket{0}{0} {}_B$ is also
divergent. 

\subsection{Complete Set of States for Complex Modes}\label{comp_state}

We start from
\begin{equation}
\bm{1} \overset{({\rm c})}{=} \sum_{n_\mu, \tilde{n}_\mu} \left[\bigotimes_\mu \kket{n_\mu,
\tilde{n}_\mu} {}_\mu \, \, {}_\mu \bbra{n_\mu, \tilde{n}_\mu} \right] \, ,
\label{eqn;compb} 
\end{equation}
with the definition 
\begin{equation}
\kket{n_\mu, \tilde{n}_\mu} {}_\mu
\equiv
\frac{1}{\sqrt{n_\mu!}\sqrt{\tilde{n}_\mu!}} (\hat{b}^\dag_\mu)^{n_\mu}
(\tilde{b}^\dag_\mu)^{\tilde{n}_\mu} \KKV_\mu \, . 
\end{equation}
Here, we defined the new symbol $\overset{({\rm c})}{=}$ which is an
equal sign only for parts belonging to the complex eigenvalues, e.g.,
$\hat{H}_0 \overset{({\rm c})}{=} \hat{H}_0^{({\rm c})}$.

Having $\hat{W}$ and $\hat{W}^{-1}$ act on Eq.~(\ref{eqn;compb})
 from the left and right, respectively, we obtain 
\begin{align}
\bm{1}
&\overset{({\rm c})}{=} \KV {}_A \,\,{}_B\BV \nn \\
&\quad {} + \sum_{\mu} \left[ \hat{B}^\dag_\mu \KV {}_A \,\,{}_B\BV
 \hat{A}_\mu \right] - \sum_{\mu} \left[ \hat{B}_\mu \KV {}_A \,\,{}_B \BV
 \hat{A}^\dag_\mu \right] \nn \\ 
& \quad {} - \sum_{\mu,\nu} \left[ \hat{B}^\dag_\mu
 \hat{B}_{\nu} \KV {}_A \,\,{}_B \BV \hat{A}^\dag_\nu \hat{A}_{\mu}
 \right]+ \sum_{\mu} \left[ \frac{1}{\sqrt{2!}}
 \hat{B}^{\dag2}_\mu \KV {}_A \,\,{} {}_B \BV \hat{A}^2_\mu
 \frac{1}{\sqrt{2!}} \right] \nn \\  
& \quad {} + \cdots \, , \label{eqn:compAB}
\end{align}
which is the complete set of states for the complex modes.
Performing the similar procedure, one obtains another representation:
\begin{align}
\bm{1}
& \overset{({\rm c})}{=} \KV {}_B \,\,{}_A\BV \nn \\
& \quad {} + \sum_{\mu} \left[ \hat{A}^\dag_\mu \KV {}_B \,\,{}_A
 \BV \hat{B}_\mu \right]  - \sum_{\mu} \left[ \hat{A}_\mu \KV {}_B \,\,{}_A\BV
 \hat{B}^\dag_\mu \right] \nn \\ 
& \quad {} - \sum_{\mu,\nu} \left[ \hat{A}^\dag_\mu
 \hat{A}_{\nu} \KV {}_B \,\,{}_A \BV \hat{B}^\dag_\nu \hat{B}_{\mu}
 \right] + \sum_{\mu} \left[ \frac{1}{\sqrt{2!}}
 \hat{A}^{\dag2}_\mu \KV {}_B \,\,{}_A\BV \hat{B}^2_{\mu}
 \frac{1}{\sqrt{2!}} \right] \nn \\  
& \quad {} + \cdots \, . \label{eqn:compBA}
\end{align}

Important observations here are that (i) the ``natural'' conjugate
of $\KV {}_A$ is ${}_B\BV$ and vice versa and that (ii) the minus sign
enters the complete set. 
The latter originates from the fact that the state space is an
indefinite metric one. 


\section{Physical States}\label{sec-Phys}
In the previous section, we studied the properties of $\hat{H}_0$ when
there exist the complex eigenvalues. 
It turned out that the state space is not the simple Fock one but the
indefinite metric one. 
Therefore, we need to impose appropriate conditions to construct a
restricted physical state space.
The physical states are required to reproduce the current experimental 
results.
It was revealed in the BEC experiment \cite{Shin} that the
condensate with a doubly quantized vortex is metastable and that its
life time is not related with the value of the imaginary part of the
complex eigenvalue of the BdG equations directly. 
Note that, in QFT, even if the system is unstable, we start with a static
picture and then the unstable behavior is described. 
Recall for example 
the description of the Beliaev process \cite{Beliaev}, that is, we
first construct the static representation with the stable unperturbed
Hamiltonian and then the decay processes are described as the higher
order of perturbation. 
Here we take the density distribution as the static quantity in accordance
with the experiment \cite{Shin}.
As discussed later, it is expected that the unstable behavior occurs due
to the external perturbation in this case.

We now require the following four \textit{physical state conditions}
(PSCs):
\def\labelenumi{(\roman{enumi})}
\begin{enumerate}
 \item $\bra{\overline{\Psi}} \hat{\psi}(x)\ket{\Psi} = \zeta (\bm{x})$, 
 \item $\bra{\overline{\Psi}} \hat{\psi}^{\dag}(x)
       \hat{\psi}(x)\ket{\Psi}$ is time-independent,
 \item $\bra{\overline{\Psi}}
       \hat{G}\ket{\Psi}$ is real, when $\hat{G}$ is an hermitian operator, 
 \item $\braket{\overline{\Psi}}{\Psi} = 1$,
\end{enumerate}
where $\ket{\Psi}$ is a physical state and $\bra{\overline{\Psi}}$ is 
its ``natural''
conjugate (see the comment below Eq.~(\ref{eqn:compBA})). 
We call $\ket{\Psi}$ and $\bra{\overline{\Psi}}$ \textit{physical 
states} if they satisfy the four conditions above.
The first and second conditions mean that the order parameter and the
density distribution are static.
The third condition ensures that the expectation value of any
hermitian operator can be interpreted as physical quantity. 
The fourth condition is necessary for the probabilistic interpretation.
Hereafter, 
we assume for simplicity that there is only one pair of complex modes,
and denote their eigenvalues, eigenfunctions and operators by
$E$ and $E^\ast$, $u ({\bm x})$, $v ({\bm x})$, $u_\ast ({\bm x})$,
and $v_\ast ({\bm x})$, and $\hat{A}$, $\hat{A}^\dag$, $\hat{B}$, and
$\hat{B}^\dag$, respectively.

From the equation $\hat{\psi} ( x) = \zeta ({\bm x}) + \hat{\varphi} (x)$,
the first condition of the PSCs (i) is equivalent to
\begin{enumerate}
 \item[(i')] $\bra{\overline{\Psi}} \hat{\varphi} (x) \ket{\Psi} = 0$.
\end{enumerate}

If $\ket{\Psi}$ and \bra{\overline{\Psi}} satisfy the condition (i'), 
the second condition (ii) states that 
\begin{enumerate}
 \item[(ii')] $\bra{\overline{\Psi}} \hat{\varphi}^{\dag}(x)
       \hat{\varphi}(x)\ket{\Psi}$ is time-independent. 
\end{enumerate}

From the discussion above, one can confirm whether a state is physical
one or not, by checking  first the condition (i'), next the condition
(ii') and (iii) , and finally the condition (iv). 
Let us examine some states,
following these steps.

\subsection{Candidate 1 --- Vacuum of $\hat{b}$ and $\tilde{b}$}

We consider the vacuum $\KKV$ as the complex part of the first
candidate for the physical state $\ket{\Psi} {}_1$, namely,
\be
\ket{\Psi}_1 \overset{({\rm c})}{=} \KKV.
\ee 
The complex part of the corresponding conjugate state is 
$\BBV$, i.e., 
\be
{}_1 \bra{\overline{\Psi}}\overset{({\rm c})}{=}\BBV.
\ee
One can easily see from the following relation: 
\begin{equation}
\BBV \hat{A} \KKV
=
\BBV \hat{A}^{\dag} \KKV
=
\BBV \hat{B} \KKV
=
\BBV \hat{A}^{\dag} \KKV
= 0 \, ,
\end{equation}
this state satisfies the first condition (i'). 

We move to the condition (ii'). It is easy to obtain
\begin{equation}
 \BBV \hat{A} \hat{A}^\dag \KKV = \BBV \hat{B} \hat{B}^\dag \KKV =
  \frac{1}{2} \, .
\end{equation}

Combining these results with 
\begin{align}
\lefteqn{\hat{\varphi}^\dag (x) \hat{\varphi} (x) } \nn \\
& \overset{({\rm c})}{=} \hat{A} \hat{A}^\dag
 \rme^{2 ({\rm Im} E) t} \left[ u ({\bm x}) u^\ast ({\bm x}) + v ({\bm
 x}) v^\ast ({\bm x}) \right] \nn \\
& \quad {} - \hat{A} \hat{A} \rme^{-2 \rmi E t} u ({\bm x}) v
 ({\bm x}) - \hat{A}^\dag \hat{A}^\dag \rme^{2 \rmi E^\ast t} u^\ast ({\bm
 x}) v^\ast ({\bm x}) \nn \\
 &\quad {} + \hat{B} \hat{B}^\dag \rme^{- 2 ({\rm Im} E) t} \left[ u_\ast
 ({\bm x}) u_\ast^\ast ({\bm x}) + v_\ast ({\bm x}) v_\ast^\ast ({\bm
 x}) \right] \nn \\
 & \quad {} - \hat{B} \hat{B} \rme^{- 2 \rmi E^\ast t} u_\ast ({\bm x})
 v_\ast ({\bm x}) - \hat{B}^\dag \hat{B}^\dag \rme^{2 \rmi E t}
 u_\ast^\ast ({\bm x}) v_\ast^\ast ({\bm x}) \nn \\
 & \quad {} - \hat{A} \hat{B} \rme^{- 2 \rmi ({\rm Re} E) t} \left[ u
 ({\bm x}) v_\ast ({\bm x}) + u_\ast ({\bm x}) v ({\bm x}) \right]
 \nn \\
 & \quad {} - \hat{A}^\dag \hat{B}^\dag \rme^{2 \rmi ({\rm Re} E) t}
 \left[ u^\ast ({\bm x}) v_\ast^\ast ({\bm x}) + u_\ast^\ast ({\bm x})
 v^\ast ({\bm x}) \right] \nn \\
 & \quad {} + \hat{A}^\dag \hat{B} u^\ast ({\bm x}) u_\ast ({\bm x})
 + \hat{B} \hat{A}^\dag v^\ast ({\bm x}) v_\ast ({\bm x}) \nn \\
 & \quad {}+  \hat{B}^\dag \hat{A} u ({\bm x}) u_\ast^\ast ({\bm x})
 + \hat{A} \hat{B}^\dag v ({\bm x}) v_\ast^\ast ({\bm x}) \, ,
 \label{vphidagvphi}
\end{align}
we can see that
$\BBV \hat{\varphi}^{\dag}(x) \hat{\varphi}(x) \KKV$ is time
dependent.
This fact means that the condition (ii') is not satisfied.
So the pair of the states $\ket{\Psi} {}_1$ and ${}_1
\bra{\overline{\Psi}}$ is excluded from physical states. 

\subsection{Candidate 2 --- Zero States}
In this subsection, we examine the zero states (see the definition
below Eq.~(\ref{eqn:0dagB})) as the second candidate for
the physical states. 

Let us pick up a pair of $\KV {}_A$ and its ``natural'' conjugate
${}_B \BV$, 
namely, we choose the complex part of the second candidate for the
physical states $\ket{\Psi}{}_2$ and ${}_2 \bra{\overline{\Psi}}$ as
\be
\ket{\Psi} {}_2 \overset{({\rm c})}{=} \ket{0}{}_A
\, , \quad
{}_2
\bra{\overline{\Psi}} \overset{({\rm c})}{=} {}_B \bra{0}
\ee
(see again the comment below Eq.~(\ref{eqn:compBA})).
We can easily check the following relations,
\begin{align}
{}_B \BV \hat{A} \KV {}_A = {}_B \BV \hat{A}^{\dag} 
\KV {}_A& = 0 \, \label{AA0} ,
 \\ 
{}_B \BV \hat{B} \KV {}_A = {}_B \BV  \hat{B}^{\dag} \KV {}_A & = 0 \label{BB0}
 \, ,
\end{align}
which means that the condition (i') is satisfied.

Next, using Eq.~(\ref{vphidagvphi}), we obtain
\begin{eqnarray}
{}_2 \bra{\overline{\Psi}} \hat{\varphi}^{\dag}(x) \hat{\varphi}(x) 
\ket{\Psi} {}_2 \overset{({\rm c})}{=}
-u^\ast (\bmx) u_\ast (\bmx)  +v(\bmx) v^\ast_\ast (\bmx) . 
\end{eqnarray}
This shows that the condition (ii') is satisfied 
but the condition (iii) is not. 

Thus we conclude that this pair of the zero states
is excluded from physical states. 
Similarly, the other pair of $\KV {}_B$ and ${}_A \BV$ 
doesn't satisfy the PSCs either.

\subsection{Candidate 3 --- Direct Sum of Zero States}
In this subsection, 
we consider the direct sum
\be
|+ \rangle 
\equiv
\frac{1}{\sqrt{2}}
\left(
\KV_A \oplus \KV_B
\right) \label{eqn:plusket}
\ee
as the complex part of the third candidate for the physical state
$\ket{\Psi} {}_3$, namely, 
\be
\ket{\Psi} {}_3 \overset{({\rm c})}{=} |+
\rangle . 
\ee
We may introduce such a direct sum 
because these two zero states belong to the different
complete sets, 
as one can see from Eqs.~(\ref{eqn:compAB}) and (\ref{eqn:compBA}). 
Note also that in this case the complex part of the corresponding
conjugate state is given as 
\be
{}_3 \bra{\overline{\Psi}} 
\overset{({\rm c})}{= }
\langle + |
\equiv
\frac{1}{\sqrt{2}}
\left(
{}_B \BV \oplus {}_A \BV
\right). \label{eqn:plusbra}
\ee 

Using Eqs.~(\ref{AA0}) and (\ref{BB0}) and their complex conjugates, 
one can see that the condition (i') is satisfied. 

Next, using Eq.~(\ref{vphidagvphi}), we obtain
\be
{}_3 \bra{\overline{\Psi}} \hat{\varphi}^{\dag} (x) \hat{\varphi} (x) 
\ket{\Psi} {}_3 \overset{({\rm c})}{=} - \mathrm{Re} \left( u^\ast
(\bmx) u_\ast (\bmx) - v(\bmx) v_\ast^\ast(\bmx) \right). 
\ee

This time, the conditions (ii') and (iii) are both satisfied. 
In particular, note the fact that the number density is time-independent. 
Thus, as already mentioned in Sec.~\ref{sec-Hamiltonian}, 
the emergence of the complex eigenmodes does not immediately imply the
instability of the system in our approach. 
It is easy to show that the condition (iv) is also satisfied. 
So, we can conclude that the third candidate for the physical states
$\ket{\Psi} {}_3$ and ${}_3 \bra{\overline{\Psi}}$ satisfy all the
PSCs. 


\section{Linear Response for Direct Sum of Zero States}\label{sec-LRT}

So far, we have developed the description of QFT in the static picture 
when the BdG equations provide complex eigenvalues. Then the complex
eigenvalues are not directly connected with the instability of the system.
If we choose the direct sum of the zero states 
as the physical states, the distribution of
the quasi-particle is stable. 
It is not yet clear how the complex modes
give rise to any instability. In order to clarify this point, 
we consider the response of the condensate against the time-dependent
external perturbations,
whose expression is derived from the linear response
theory \cite{LRTtext,Minguzzi,Kubo}. 
It will turn out that
the response is remarkably affected by the presence of complex modes.

We consider the following external perturbation which corresponds to
time-dependent modification of the trap:
\begin{equation}
 \hat{H}_{\mathrm{ex}} (t) = \int \! d^3 x \, \hat{\psi}^{\dag}(x)
\delta V_{\mathrm{ex}}(x) \hat{\psi}(x) \, .
\end{equation}

From the linear response theory, the response of the condensate density
distribution $\delta \langle \hat{\rho}(\bmx, t) \rangle$ is given as
\begin{equation}
\delta \langle \hat{\rho}(x) \rangle
=
\int \! d^4 x' \, G_{\mathrm{R}}(\bmx, \bmx', t-t') \delta
V_{\mathrm{ex}}(x')  \, ,
\end{equation}
where
\begin{equation}
\hat{\rho}(x) = \hat{\psi}^{\dag}(x)\hat{\psi}(x)
\end{equation}
and $\delta V_{\mathrm{ex}}(x)$ is a
time-dependent external potential.  The retarded Green function
$G_{\mathrm{R}}(\bmx, \bmx', t-t')$ is expressed as
\begin{align}
G_{\mathrm{R}}(\bmx, \bmx', t-t') &= -i \theta(t-t')
\langle [ \hat{\psi}^{\dag}(x) \hat{\psi}(x) , \,
 \hat{\psi}^{\dag}(x') \hat{\psi}(x') ] \rangle \nn \\
& =
- i \zeta^\ast (\bmx) \zeta(\bmx') \theta(t-t')
\langle \hat{\varphi}(\bmx, t) \hat{\varphi}^{\dag} (\bmx', t') \rangle \nn \\
& \quad {}
+ i \zeta ^\ast (\bmx) \zeta (\bmx') \theta(t-t')
\langle \hat{\varphi}^{\dag} (\bmx', t') \hat{\varphi}(\bmx, t) \rangle \nn \\
& \quad {}
- i \zeta (\bmx) \zeta^\ast(\bmx') \theta(t-t')
\langle \hat{\varphi}^{\dag} (\bmx, t) \hat{\varphi}(\bmx', t') \rangle \nn \\
& \quad {}
+ i \zeta (\bmx) \zeta^\ast(\bmx') \theta(t-t')
\langle \hat{\varphi}(\bmx', t') \hat{\varphi}^{\dag} (\bmx, t) \rangle \nn \\
& \quad {}
+ \cdots \, ,
\label{respgen}
\end{align}
where the symbol ``$\cdots$'' stands for the higher order terms of
$\hat{\varphi}$ and $\hat{\varphi}^{\dag}$,  and $\langle \cdot \rangle$
represents the expectation value of some physical states.
In the following discussion, we consider the case in which there is a
single pair of complex modes as in the previous section, 
and employ the states $\ket{\Psi} {}_3$ and $_3 \bra{\overline{\Psi}}$ as
the physical states, whose complex parts are given as $\ket{\Psi} {}_3
\overset{({\rm c})}{=} \ket{+}$ and $_3 \bra{\overline{\Psi}}
\overset{({\rm c})}{=} \bra{+}$ (see Eqs.~(\ref{eqn:plusket})--(\ref{eqn:plusbra})).

The type of the time-dependent external potential should correspond to
an experimental setup, and we choose two types: One is an impulsive force, 
and the other is an oscillating one.
In the case of an oscillating external perturbation, we assume that the
perturbation starts at $t_0$ and is kept through.

First, we focus on the impulsive external potential.
In this case, the external potential is given as
\begin{equation}
 \delta V_{\mathrm{ex}} (x) = \delta(t-t_0) \delta \tilde{V}(\bmx)
  \, .
\end{equation}
From Eq.~(\ref{respgen}), one obtains the following density response,
\begin{align}
 \delta \langle \hat{\rho}(x) \rangle
&=
- i \zeta^\ast(\bmx) u(\bmx)
\left( \int \! d^3 x' \, \zeta (\bmx') u_\ast^\ast(\bmx') \delta
 \tilde{V}(\bmx') \right) e^{-i E (t-t_0)} \nn \\
& \quad {}
- i \zeta^\ast(\bmx) u_\ast(\bmx)
\left(
\int \! d^3 x' \, \zeta(\bmx') u^\ast(\bmx') \delta \tilde{V}(\bmx')
\right) e^{-i E^\ast (t-t_0)} \nn \\
& \quad {}
+ i \zeta^\ast(\bmx) v^\ast(\bmx)
\left( \int \! d^3 x' \, \zeta(\bmx') v_\ast(\bmx') \delta
\tilde{V}(\bmx') \right) e^{i E^\ast (t-t_0)} \nn \\
& \quad {}
+
i \zeta^\ast(\bmx) v_\ast^\ast(\bmx)
\left( \int \! d^3 x' \, \zeta(\bmx') v(\bmx') \delta \tilde{V}(\bmx')
\right) e^{i E (t-t_0)} \nn \\
& \quad {}
- i \zeta(\bmx) v(\bmx)
\left( \int \! d^3 x' \, \zeta^\ast(\bmx') v_\ast^\ast(\bmx') \delta
 \tilde{V}(\bmx') \right) e^{-i E (t-t_0)} \nn \\
& \quad {}
- i \zeta(\bmx) v_\ast(\bmx)
\left( \int \! d^3 x' \, \zeta^\ast(\bmx') v^\ast(\bmx') \delta
\tilde{V}(\bmx') \right) e^{-i E^\ast (t-t_0)} \nn \\
& \quad {}
+ i \zeta(\bmx) u^\ast(\bmx)
\left( \int \! d^3 x' \, \zeta^\ast(\bmx') u_\ast(\bmx') \delta
 \tilde{V}(\bmx') \right) e^{i E^\ast (t-t_0)} \nn \\
& \quad {}
+
i \zeta(\bmx) u_\ast^\ast(\bmx)
\left( \int \! d^3 x' \, \zeta^\ast(\bmx') u(\bmx') \delta
\tilde{V}(\bmx') \right) e^{i E (t-t_0)} \, .
\end{align}

It is seen from the above expression that  
the density response $\delta\langle \hat{\rho}(x)
\rangle$  blows up or damps in general if the complex modes exist. 
In this sense, we can say that the complex eigenmodes cause the
instability against the external perturbation.  
This result is consistent with that of the analyses using the TDGP
equation \cite{Mettenen,Kawaguchi}. 

Next, apply an oscillating external potential which is given as
\begin{equation}
 \delta V_{\mathrm{ex}} (x) = \theta(t-t_0) R (\bmx) \cos
  \tilde{\omega} t \, .
\end{equation}
Here $R(\bmx)$ is some function of $\bmx$ representing a type
of oscillation. For example, one may choose $R({\bm x}) = x$ in the case
of the dipole oscillation of the condensate to the $x$-direction.
The general expression of the response of the condensate is given as
\begin{align}
 \delta \langle \hat{\rho}(x) \rangle
&=
- 
\frac{1}{2}
\frac{K \left[ \zeta, u_\ast^\ast \right]}{\tilde{\omega} + E}
\zeta^\ast(\bmx) u(\bmx) e^{i \tilde{\omega} t_0}
\left( e^{i \tilde{\omega} (t-t_0)} - e^{- i E (t-t_0)} \right) \nn
 \\
& \quad {}+
\frac{1}{2}
\frac{K \left[ \zeta, u_\ast^\ast \right]}{\tilde{\omega} - E}
\zeta^\ast(\bmx) u(\bmx) e^{-i \tilde{\omega} t_0}
\left( e^{-i \tilde{\omega} (t-t_0)} - e^{- i E (t-t_0)} \right)
 \nn \\
& \quad {}
- \frac{1}{2}
\frac{K \left[ \zeta, u^\ast \right]}{\tilde{\omega} + E^\ast }
\zeta^\ast(\bmx) u_\ast(\bmx) e^{i \tilde{\omega} t_0}
\left(
e^{i \tilde{\omega} (t-t_0)} - e^{- i E^\ast (t-t_0)}
\right) \nn \\
& \quad {} +
\frac{1}{2}
\frac{K \left[\zeta, u^\ast \right]}{\tilde{\omega} - E^\ast}
\zeta^\ast(\bmx) u_\ast(\bmx) e^{-i \tilde{\omega} t_0}
\left(
e^{-i \tilde{\omega} (t-t_0)} - e^{- i E^\ast (t-t_0)}
\right) \nn \\
& \quad {}
+
\frac{1}{2}
\frac{K \left[\zeta, v_\ast \right]}{\tilde{\omega} - E^\ast}
\zeta^\ast(\bmx) v^\ast(\bmx) e^{i \tilde{\omega} t_0}
\left( e^{i \tilde{\omega} (t-t_0)} - e^{ i E^\ast (t-t_0)} \right) \nn \\
& \quad {} -
\frac{1}{2}
\frac{K \left[\zeta, v_\ast \right]}{\tilde{\omega} + E^\ast}
\zeta^\ast(\bmx) v^\ast(\bmx) e^{-i \tilde{\omega} t_0}
\left( e^{-i \tilde{\omega} (t-t_0)} - e^{ i E^\ast (t-t_0)} \right) \nn \\
& \quad {}
- \frac{1}{2}
\frac{K \left[\zeta, v \right]}{\tilde{\omega} - E}
\zeta^\ast(\bmx) v_\ast^\ast(\bmx) e^{i \tilde{\omega} t_0}
\left( e^{i \tilde{\omega} (t-t_0)} - e^{ i E (t-t_0)} \right) \nn
\\
& \quad {} +
\frac{1}{2}
\frac{K \left[ \zeta, v \right]}{\tilde{\omega} + E}
\zeta^\ast(\bmx) v_\ast^\ast(\bmx) e^{-i \tilde{\omega} t_0}
\left( e^{-i \tilde{\omega} (t-t_0)} - e^{ i E (t-t_0)} \right) \nn
 \\
& \quad {}
- \frac{1}{2}
\frac{K \left[ \zeta^\ast, v_\ast^\ast \right]}{\tilde{\omega} + E}
\zeta(\bmx) v(\bmx) e^{i \tilde{\omega} t_0}
\left( e^{i \tilde{\omega} (t-t_0)} - e^{- i E (t-t_0)} \right) \nn
 \\
& \quad {} +
\frac{1}{2}
\frac{K \left[ \zeta^\ast, v_\ast^\ast \right]}{\tilde{\omega} - E}
\zeta(\bmx) v(\bmx) e^{-i \tilde{\omega} t_0}
\left( e^{-i \tilde{\omega} (t-t_0)} - e^{- i E (t-t_0)} \right)
 \nn \\
& \quad {}
- \frac{1}{2}
\frac{K \left[ \zeta^\ast, v^\ast \right]}{\tilde{\omega} + E^\ast}
\zeta (\bmx) v_\ast(\bmx) e^{i \tilde{\omega} t_0}
\left( e^{i \tilde{\omega} (t-t_0)} - e^{- i E^\ast (t-t_0)} \right) \nn \\
& \quad {} +
\frac{1}{2}
\frac{K \left[ \zeta^\ast, v^\ast \right]}{\tilde{\omega} - E^\ast}
\zeta(\bmx) v_\ast(\bmx) e^{-i \tilde{\omega} t_0}
\left( e^{-i \tilde{\omega} (t-t_0)} - e^{- i E^\ast (t-t_0)} \right) \nn \\
& \quad {}
+ \frac{1}{2}
\frac{K \left[ \zeta^\ast, u_\ast \right]}{\tilde{\omega} - E^\ast}
\zeta(\bmx) u^\ast(\bmx) e^{i \tilde{\omega} t_0}
\left( e^{i \tilde{\omega} (t-t_0)} - e^{ i E^\ast (t-t_0)}
\right) \nn \\
& \quad {} -
\frac{1}{2}
\frac{K \left[ \zeta^\ast, u_\ast \right]}{\tilde{\omega} + E^\ast}
\zeta (\bmx) u^\ast(\bmx) e^{-i \tilde{\omega} t_0}
\left( e^{-i \tilde{\omega} (t-t_0)} - e^{ i E^\ast (t-t_0)} \right) \nn \\
&\quad {}
+
\frac{1}{2}
\frac{K \left[ \zeta^\ast, u \right]}{\tilde{\omega} - E}
\zeta(\bmx) u_\ast^\ast(\bmx) e^{i \tilde{\omega} t_0}
\left( e^{i \tilde{\omega} (t-t_0)} - e^{ i E (t-t_0)} \right) \nn
 \\
& \quad {} -
\frac{1}{2}
\frac{K \left[ \zeta^\ast, u \right]}{\tilde{\omega} + E}
\zeta(\bmx) u_\ast^\ast(\bmx) e^{-i \tilde{\omega} t_0}
\left( e^{-i \tilde{\omega} (t-t_0)} - e^{ i E (t-t_0)} \right) \, ,
\end{align}
where
\begin{align}
K \left[ \zeta , u_\ast^\ast \right] &= \int \! d^3 x' \, R(\bmx')
 \zeta(\bmx') u_\ast^\ast(\bmx') \, , \\
K \left[ \zeta^\ast , v \right] &= \int \! d^3 x' \, R(\bmx')
 \zeta^\ast(\bmx') v(\bmx') \, .
\end{align}

This again shows that the instability against
the external perturbation comes from the complex eigenmodes.

\section{Summary}\label{sec-Sum}
In this paper, we have 
constructed the description of QFT (quantum field theory) starting with
a static picture and have formulated Kubo's LRT (linear response
theory) in the case where complex eigenmodes of the BdG equations arise.  
First, we have summarized mathematical properties of the eigenfunctions. 
These properties guide us to a consistent form of the
complete set.
Using the complete set, we expand quantum fields and obtain the
representation of the free Hamiltonian. 
We have shown that the Hamiltonian is not diagonalizable in the
conventional bosonic representation. 
The eigenstates of the free Hamiltonian can be constructed, although
their ``norms'' diverge. These intricate circumstances, when the complex
eigenmodes are involved, come from the fact that an indefinite metric
space comes in. To deal with an indefinite metric and to construct a
physical state space, we have introduced the PSCs (physical state
conditions), reflecting the metastability of doubly quantized vortex
states of BEC.
We investigate the candidates for the physical states.
It turns out that if we employ the vacuum of $\hat{b}$ and
$\tilde{b}$ (Candidate 1) and the zero states $\KV_A$ or $\KV_B$
(Candidate 2) as the complex part of the states, they don't satisfy all
the PSCs. 
But, if we choose the direct sum of the zero states (Candidate 3) as
the complex part of the states, they satisfy the PSCs. 
Using the direct sum of the zero states as the complex part of the
physical states, we can start with the stable representation in QFT and
calculate the (retarded) Green function.  
Finally the density response against the external perturbations (the
impulse and oscillating types) is derived.
This result shows blow-up and damping behavior of the fluctuations,
which is qualitatively consistent with the result of the analyses using the TDGP
equations \cite{Mettenen,Kawaguchi}. 
Thus the complex eigenmodes cause the instability against the external
perturbation. 

We mention that the higher order terms of $\hat{\varphi}$ and
$\hat{\varphi}^{\dag}$ have contributions to the density response, 
and have the poles at $E_n \pm E$ and so on, which are the contribution
beyond the calculations based on the TDGP equation.



\section*{Acknowledgments}
The authors would like to thank Professor I.~Ohba and Professor
 H.~Nakazato for helpful comments and encouragement, 
and Dr.~M.~Miyamoto and Dr.~K.~Kobayashi for useful discussions.  
M.M. and T.S. are supported partially by the Grant-in-Aid for The 21st
Century COE Program (Physics of Self-organization Systems) at Waseda
University.
This work is partly supported by a Grant-in-Aid for Scientific Research
(C) (No.~17540364) from the Japan Society for the Promotion of Science,
for Young Scientists (B) (No.~17740258) and for
Priority Area Research (B) (No.~13135221) both from the Ministry of
Education, Culture, Sports, Science and Technology, Japan.


\end{document}